\documentclass[preprint,nofootinbib]{revtex4}%
\usepackage{amssymb}
\usepackage{amsfonts}
\usepackage{amsmath}
\usepackage[utf8]{inputenc}
\usepackage{fancyhdr}
\usepackage{amsmath}
\usepackage{graphicx}
\usepackage{hyperref}
\usepackage{subcaption}
\usepackage{url}
\usepackage[usenames]{color}%
\providecommand{\U}[1]{\protect\rule{.1in}{.1in}}
\providecommand{\U}[1]{\protect\rule{.1in}{.1in}}
\definecolor{blue}{rgb}{0,0,1}

\definecolor{red}{rgb}{1,0,0}

\begin{document}
\title{Slowly rotating black holes in Quasi-topological gravity}
\author{Octavio Fierro$^{1}$, Nicol\'{a}s Mora$^{2}$, Julio Oliva$^{2}$}
\affiliation{$^{(1)}$Departamento de Matem\'{a}tica y F\'{\i}sica Aplicadas, Universidad
Cat\'{o}lica de la Sant\'{\i}sima Concepci\'{o}n, Alonso de Ribera 2850,
Concepci\'{o}n, Chile.}
\affiliation{$^{(2)}$Departamento de F\'{\i}sica, Universidad de Concepci\'{o}n, Casilla,
160-C, Concepci\'{o}n, Chile.}

\begin{abstract}
While cubic Quasi-topological gravity is unique, there is a family of quartic
Quasi-topological gravities in five dimensions. These theories are defined by
leading to a first order equation on spherically symmetric spacetimes, resembling the structure of the equations of Lovelock theories in higher-dimensions, and are also ghost free around
AdS. Here we construct slowly rotating black holes in these theories, and show
that the equations for the off-diagonal components of the metric in the cubic
theory are automatically of second order, while imposing this as a restriction
on the quartic theories allows to partially remove the degeneracy of these theories, leading to a three-parameter family of Lagrangians of order four in the Riemann
tensor. This shows that the parallel with Lovelock theory observed on
spherical symmetry, extends to the realm of slowly rotating solutions. In the
quartic case, the equations for the slowly rotating black hole are obtained
from a consistent, reduced action principle. These functions admit a simple
integration in terms of quadratures.  Finally, in order to go beyond the slowly rotating regime, we explore the consistency of the Kerr-Schild ansatz in cubic Quasi-topological gravity. Requiring the spacetime to asymptotically match with the rotating black hole in GR, for equal oblateness parameters, the Kerr-Schild deformation of an AdS vacuum, does not lead to a solution for generic values of the coupling. This result suggests that in order to have solutions with finite rotation in Quasi-topological gravity, one must go beyond the Kerr-Schild ansatz.

\end{abstract}
\maketitle

\section{Introduction}

In order to construct a gravitational Lagrangian beyond the Einstein-Hilbert
combination, leading to second order field equations for the metric tensor of
a generic spacetime, one is forced to go beyond four dimensions. Lovelock
theory emerge as the general, diffeomorphism invariant gravitational action
that is constructed with the metric as the unique field and leads to second
order field equations \cite{Lovelock:1971yv}. Each Lovelock term has a simple
interpretation as the dimensional continuation of a lower dimensional Euler
density \cite{Zumino:1985dp}, and have received considerable attention during the last
decades since they permit to study in a controlled framework, the effects of
higher curvature terms in the gravitational dynamics (see e.g.
\cite{GastonCecilia}, \cite{CharmousisReview}, \cite{MyersGB} and references therein). The Einstein-Gauss-Bonnet Lagrangian, which is the first Lovelock theory beyond GR, also appears in the low-energy limit of Heterotic String Theory \cite{GBfromStrings}. In such setup, the
higher curvature terms are consistent only in a perturbative regime
\cite{Camanho:2014apa}, an interpretation that is supported by studies of the
initial value problem in this framework see e.g. \cite{Willison:2014era}, \cite{Papallo:2017qvl} and \cite{Andrade:2016yzc}. In this scenario, going beyond
quadratic corrections may lead to higher curvature combinations outside of the
Lovelock family, even after using the field redefinitions proper of the
perturbative approach \cite{Metsaev:1986yb}. Furthermore the number of independent,
non-trivial Lovelock terms is bounded from above by $n<D/2$, where $n$ is the
degree of homogeneity of the Lovelock combination that goes as $Riem^{n}$. In
this manner, in a five-dimensional holographic setup and within the realm of
Lovelock gravities one cannot go beyond $Riem^{2}$ terms. Generic higher
curvature terms will propagate unhealthy degrees of freedom even around the
maximally symmetric vacuum of the theory, and in such setups it is usually
impossible to construct analytic, spherically symmetric black holes.

In \cite{Oliva:2010eb} a new five-dimensional combination of cubic terms was originally
introduced, which has interesting properties: 1) it leads to second order
field equations on a generic spherically symmetric spacetime, 2) it possesses
a Birkhoff's theorem, namely spherical symmetry implies staticity for generic
values of the couplings, 3) It possesses asymptotically flat and (A)dS black
holes that are determined by the solution of an algebraic cubic equation,
mimicking the structure of cubic Lovelock theories in dimension greater than
six. The authors of \cite{Myers:2010ru} arrived to the same cubic combination in dimension
five and dubbed it Quasi-topological gravity (a terminology that hereafter we
adopt). They found that even though the field equations are of fourth order
for generic backgrounds, the perturbations around an AdS vacuum of the theory
are of second order. Actually the equation for the graviton on AdS in
Quasi-topological gravity reduces to that in GR, with an effective Newton's
constant. In the original reference \cite{Oliva:2010eb} it was actually recognized that the cubic
combination was a particular case of a family of Lagrangians containing terms
of the form $Riem^{k}$ in dimension $D=2k-1$, which can be written as%
\begin{align}
\mathcal{\tilde{L}}_{k}={\frac{1}{2^{k}}}\left(  \frac{1}{D-2k+1}\right)
\delta_{c_{1}d_{1}\cdots c_{k}d_{k}}^{a_{1}b_{1}\cdots a_{k}b_{k}}\left(
C_{a_{1}b_{1}}^{c_{1}d_{1}}\cdots C_{a_{k}b_{k}}^{c_{k}d_{k}}-R_{a_{1}b_{1}%
}^{c_{1}d_{1}}\cdots R_{a_{k}b_{k}}^{c_{k}d_{k}}\right)  \nonumber\\
-c_{k}C_{a_{1}b_{1}}^{a_{k}b_{k}}C_{a_{2}b_{2}}^{a_{1}b_{1}}\cdots
C_{a_{k}b_{k}}^{a_{k-1}b_{k-1}}\ .\label{Lk}%
\end{align}
Here $C_{a\ cd}^{\ b}$ is the Weyl tensor and
\begin{equation}
c_{k}={\frac{(D-4)!}{(D-2k+1)!}}{\frac
{[k(k-2)D(D-3)+k(k+1)(D-3)+(D-2k)(D-2k-1)]}{[(D-3)^{k-1}(D-2)^{k-1}%
+2^{k-1}-2(3-D)^{k-1}]}\ .}%
\end{equation}

The cubic Quasi-topological theory is therefore obtained by setting $D=5$ and
$k=3$ in Eq. (\ref{Lk}) once the generalized Kronecker delta has been
expanded. Notice that no singular limit has to be taken in order to make sense
of the cubic combination (\ref{Lk}) in $D=5$, since after the generalized
Kronecker delta is expanded, the would-be singular factor is cancelled.

The main objective of the present paper is to fill a gap and to study rotating
solutions of Quasi-topological gravities. In the presence of higher curvature
terms, rotating solutions are notoriously difficult to study and often one is
forced to rely on approximate and/or numerical schemes. For example even in the simplest
Lovelock theory beyond GR, the Einstein-Gauss-Bonnet theory, for generic
values of the couplings, the rotating black hole solutions are only known
numerically \cite{Brihaye:2008kh}, \cite{Konoplya:2020fbx} and perturbative in the angular momentum
\cite{Kim:2007iw,Yue:2011et}. Analytic, rotating solutions can be
constructed for the special values of couplings that allows formulating the
theory as Chern-Simons theory for the AdS group \cite{Anabalon:2009kq} (see also \cite{Cvetic:2016sow}). The exact solution in five dimensions is
non-circular \cite{Anabalon:2010ns} and may have an intricate causal structure.

In five dimensions, one can also construct quartic Quasi-topological
combinations. The first of these combinations was presented in \cite{Dehghani:2011vu}. At the level of spherical symmetry, the cubic theory in
five dimensions is unique, while there is a family of quartic theories leading
to second order field equations. A family of quintic Quasi-topological
theories was also constructed in \cite{Cisterna:2017umf}.
Surprisingly in \cite{Bueno:2019ycr} it has been recently realized that a
recursive method exists which permits to construct Quasi-topological theories
of arbitrary order.

\bigskip

In this paper we construct slowly rotating solutions of cubic and quartic
Quasi-topological theories. We show that requiring
the equations for the rotating function profile to be of second order, further
restricts the couplings of the quartic combinations. We also provide a no-go results about the usefulness of the Kerr-Schild ansatz beyond the slowly rotating regime.

\bigskip

This paper is organized as follows: In Section II we review the main
properties of Cubic Quasi-topological gravity in what regards to the structure of
vacua and static black holes. In Section III, in order to approach the problem
of the rotating solutions in Cubic Quasi-topological gravity we construct the
slowly rotating black holes with two angular momenta. The functions
controlling the rotation fulfill a second order system that can be integrated
in terms of quadratures. Section IV contains an extension of these results for
quartic Quasi-topological theories in five dimensions, and remarkably we show
that requiring a second order equation for the off-diagonal terms allows to
further constrain the coefficients of the quartic terms, leading to four
different families of theories. As a first attempt to go beyond the slowly
rotating regime, we explore the Kerr-Schild ansatz in Section V. For the cubic
case we show that it is impossible to accommodate a rotating solution on this
ansatz for generic values of the couplings, requiring the leading behavior at
infinity to be that of GR, as it occurs on spherical symmetry. After proving
this no-go result in Section VII we conclude and provide further comments.

\section{Reviewing the vacua and static black holes in Cubic Quasi-topological
gravity}

We will work with the theory%
\begin{align}
I_{CQTG}&= \int \sqrt{-g}d^5x\bigg(R-2\Lambda_0+\alpha_2\left(R^2-4R_{ab}R^{ab}+R_{abcd}R^{abcd}\right)\nonumber\\
&\qquad+\alpha_3\left(R_{abcd}R^{bedf}R_{e\;f}^{\;\,a\;\,c}-\frac{9}{7}R_{abcd}R^{abce}R^d_{\,e}+\frac{3}{8}R_{abcd}R^{abcd}R\right.\nonumber\\
&\quad\;\;\qquad\left.+\frac{15}{7}R_{abcd}R^{ac}R^{bd}+\frac{18}{7}R_a^bR_b^cR_c^a-\frac{33}{14}R_{ab}R^{ab}R+\frac{15}{56}R^3\right)\bigg)\ .
\label{LaCubica}%
\end{align}

The metric%
\begin{equation}
ds^{2}=-f\left(  t,r\right)  dt^{2}+\frac{dr^{2}}{g\left(  t,r\right)  }%
+r^{2}d\Sigma_{3}^{2}\ ,
\end{equation}
is the most general ansatz for a spacetime that is compatible with the
possible isometries of the three-dimensional Euclidean manifold $\Sigma_{3}$,
which we assume of constant curvature $\gamma=\pm1,0$. As shown in \cite{Oliva:2010eb}, the
field equations of Quasi-topological gravity imply, up to a redefinition of
the coordinate $t$, that $g\left(  t,r\right)  =f\left(  t,r\right)  =f\left(
r\right)  $ where the function $f\left(  r\right)  $ is a solution of the
following cubic polynomial equation%
\begin{equation}
24\alpha_3\left(  f-\gamma\right)  ^{3}+84\alpha_2 r^{2}\left(  f-\gamma\right)
^{2}-42r^{4}\left(  f-\gamma\right)  -7r^{6}\Lambda_{0}=42mr^{2}%
\ ,\label{polyf}%
\end{equation}
where $m$ is an integration constant that will determine the energy content of
the spacetime. These black holes can also be dressed by a conformal scalar hair \cite{Chernicoff:2016qrc,Dykaar:2017mba}. The curvature radii $l$, of the maximally
symmetric (A)dS solutions of the theory are determined from (\ref{polyf})
setting $m=0$ and $f=\gamma-\Lambda_{eff}r^{2}/6$, leading to%
\begin{equation}
\Upsilon\left[  \Lambda_{eff}\right]  :=\alpha_3\Lambda_{eff}^{3}-21\alpha_2
\Lambda_{eff}^{2}-63\Lambda_{eff}+63\Lambda_{0}=0\ .\label{polyl}%
\end{equation}

Unless otherwise stated we will assume that the couplings of the theory are
generic, which implies that there is always a solutions of (\ref{polyl}) that
goes to $\Lambda_{eff}^{\left(  i\right)  }\rightarrow\Lambda_{0}$, as the
couplings of the quadratic and cubic terms in the action vanish\footnote{Two
of these vacua coincide provided the couplings fulfill the constraint
$\alpha_3^{2}\Lambda_{0}^{2}-14\left(  \alpha_2\Lambda+\frac{1}{3}\right)
\alpha_3-\frac{49}{9}\alpha_2^{2}(4\alpha_2\Lambda_{0}+3)=0$, while the three of
them degenerate provided $\alpha_2\Lambda_{0}=-1$ and $\alpha_3=-\frac{7}{3\Lambda_0}$, leading to a single cosmological
constant $\Lambda=3\Lambda_{0}$. These case are incompatible with the
interpretation of the higher curvature terms as perturbative corrections to
GR, but may allow to enlarge the space of solutions or even enlarge the
symmetry group of the theory as it happens in Lovelock theories \cite{libro}.}. Here $i\in\left\{  1,2,3\right\}  $. The three possible
solutions of (\ref{polyf}) approach each of the three possible asymptotic
curvatures $\Lambda_{eff}^{\left(  i\right)  }$. Namely the three solutions of
(\ref{polyf}), behave asymptotically as%
\begin{equation}
f^{\left(  i\right)  }\left(  r\right)  =-\frac{\Lambda_{eff}^{\left(
i\right)  }}{6}r^{2}+\gamma-\frac{m}{r^{2}}+O\left(  r^{-3}\right)  \ ,
\end{equation}
and as usual we define the GR branch as the one that smoothly connects with
the Einstein solution as%
\begin{equation}
\lim_{\left(  \alpha_3,\alpha\right)  \rightarrow0}\Lambda_{eff}^{\left(  i\right)
}\left(  \kappa,\alpha,\Lambda_{0}\right)  \rightarrow\Lambda_{0}\ .
\end{equation}
Hereafter we will focus on this branch.

\begin{figure}[!ht]
     \centering
     \begin{subfigure}[b]{0.42\textwidth}
         \centering
         \includegraphics[width=\textwidth]{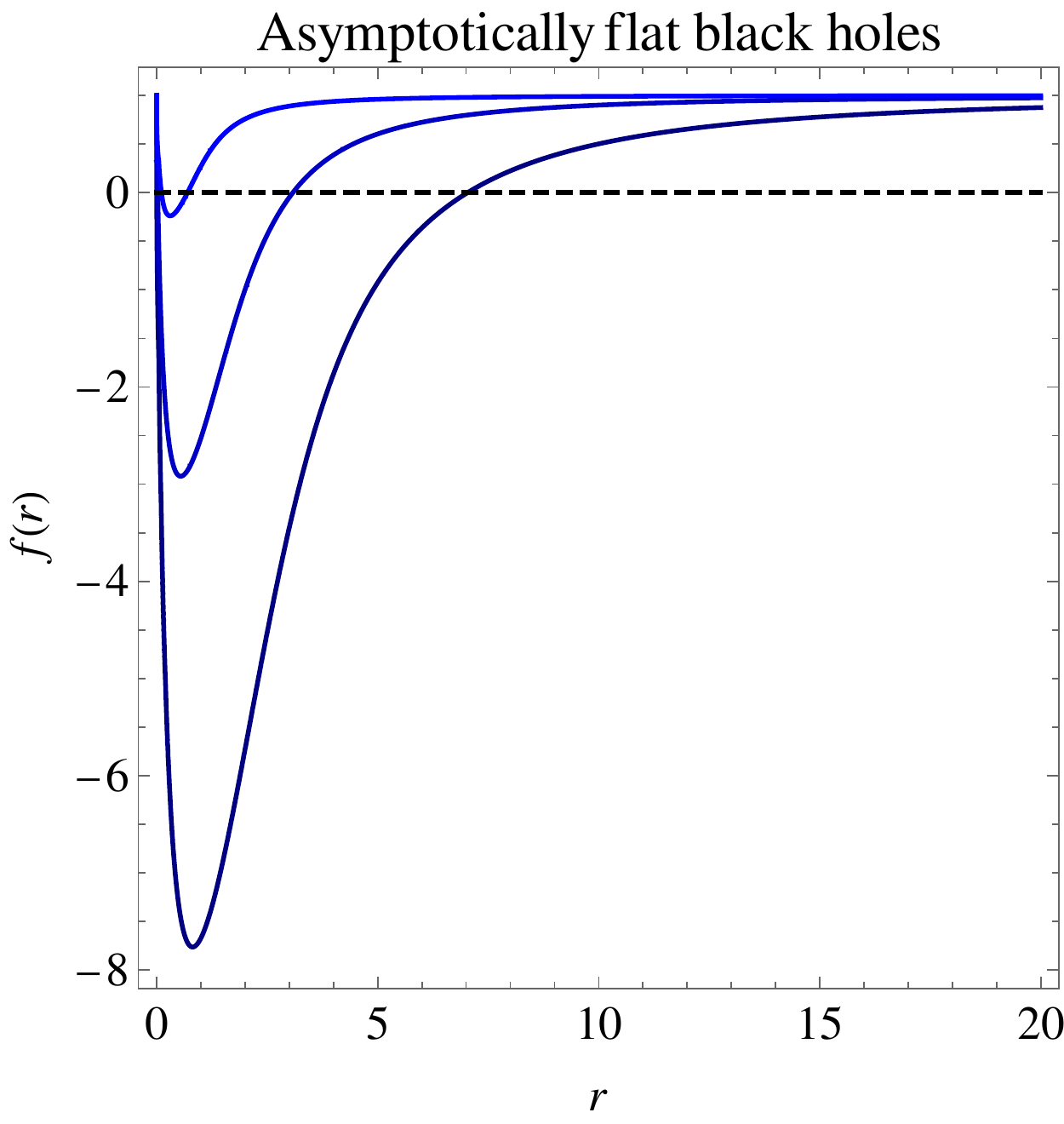}
     \end{subfigure}
     \hfill
     \begin{subfigure}[b]{0.42\textwidth}
         \centering
         \includegraphics[width=\textwidth]{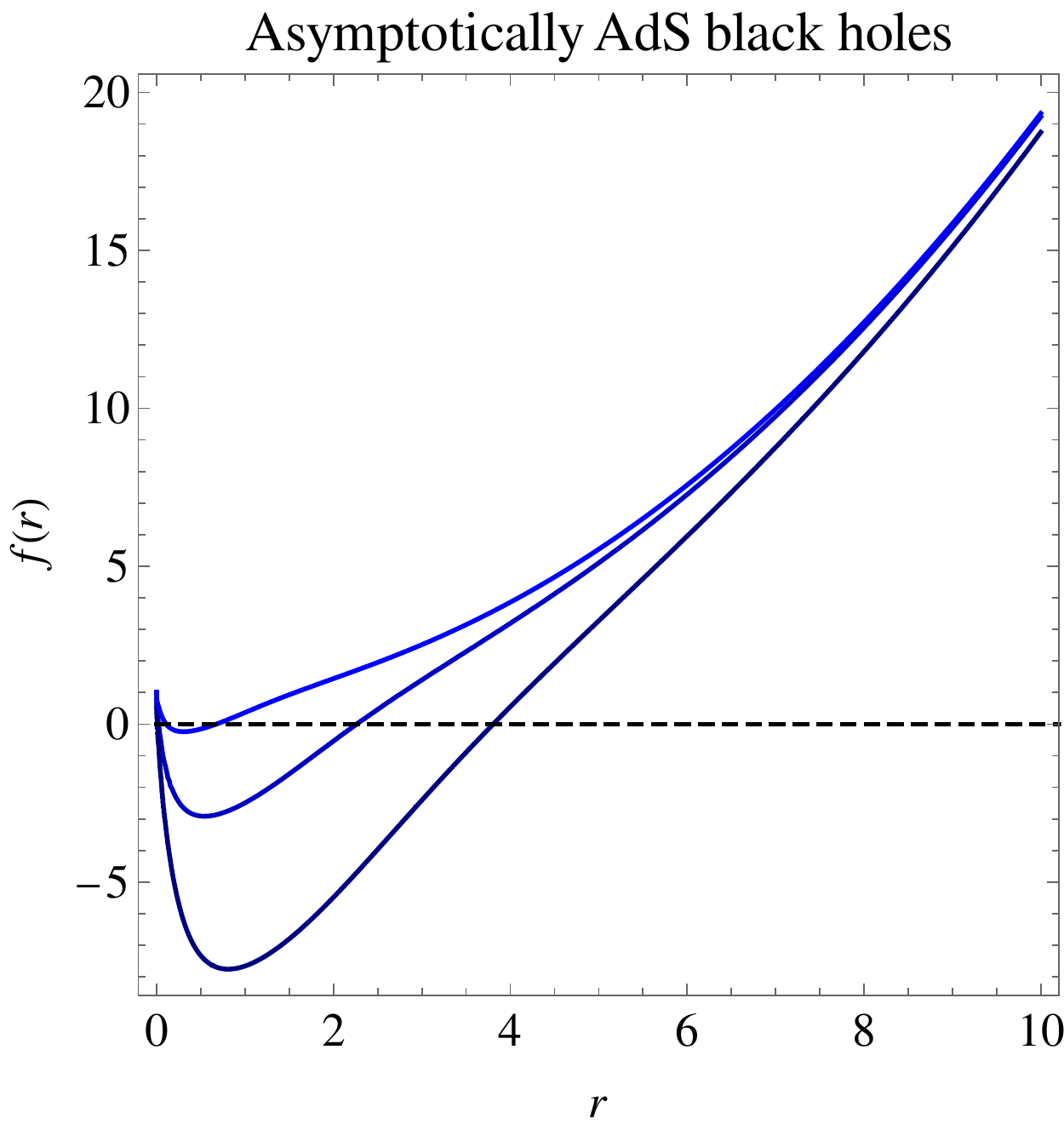}
     \end{subfigure}
    \caption{The figure depicts the function $f(r)$  for
asymptotically flat black holes (left panel) and asymptotically AdS black
holes (right panel) for different values of the mass parameter $m$ in Eq.
(\ref{polyf}), in particular $m=1, 10, 50$. The profiles were obtained for small values of the higher
curvature couplings $\alpha_2=0.25$ and $\alpha_3=-0.01$, while for the
asymptotically AdS black holes the bared cosmological term $\Lambda_{0}$ was
set equal to $-1$}
        \label{AsymBH}
\end{figure}

Even though the solutions of the cubic equation can be written explicitly in
terms of radicals, they are not particularly illuminating and therefore we
only present some examples in Figure \ref{AsymBH}.

As observed in \cite{Oliva:2010eb}, the polynomial (\ref{polyf}) resembles the structure of
Wheeler's polynomial in cubic Lovelock theory \cite{Wheeler:1985nh,Wheeler:1985qd}, formally replacing
$D=5$ after a $\left(  D-5\right)  $ factor is absorbed in the coupling of the
cubic Lovelock terms (see also Eq. (5.8) of \cite{Maeda:2011ii} as well as \cite{Camanho:2009hu,Camanho:2011rj,Corral:2019leh}). The same pattern extends to Quasi-topological
gravities of higher degree (\cite{Cisterna:2017umf,Bueno:2019ycr}).

\bigskip

In the next section we move forward and construct the slowly rotating black
hole solutions of Cubic Quasi-topological gravity.

\section{Slowly rotating black holes in Quasi-topological gravity}

The metric ansatz that accommodates the slowly rotating black hole solution of
Quasi-topological gravity is%

\begin{align}
ds^{2} &  =-f\left(  r\right)  dt^{2}+\frac{dr^{2}}{f\left(  r\right)  }%
+r^{2}\left(  \frac{d\mu^{2}}{1-\mu^{2}}+\mu d\phi^{2}+\left(  1-\mu\right)
^{2}d\psi^{2}\right)  \label{slowly}\nonumber\\
& \;\quad -a_{1}\mu^{2}r^{2}g_{1}\left(  t,r\right)  dtd\phi-a_{2}\left(  1-\mu
^{2}\right)  r^{2}g_{2}\left(  t,r\right)  dtd\psi\ ,
\end{align}
with $-\infty<t<\infty$, $-1\leq\mu\leq1$, and $0\leq\phi\leq2\pi$, $0\leq
\psi\leq2\pi$. Here $a_{1}$ and $a_{2}$ are the rotation parameters which we
will consider small. Notice that we have chosen the static metric to be
spherically symmetric, even though the present analysis can be extended for
arbitrary curvature of the horizon of the static black hole, which may lead to
rotating, topological black branes as in GR in four dimensions \cite{Klemm:1997ea}. As
usual, expanding the field equations in powers of $a_{1}$ and $a_{2}$ leads,
to the lowest order, to the same equation for $f\left(  r\right)  $ that
implies (\ref{polyf}), and we write it here for future purposes in its
differential form%
\begin{equation}
\left(  \frac{12\left(  f-1\right)  ^{2}\alpha_3}{7r^{2}}+4\left(  f-1\right)
\alpha_2-r^{2}\right) \frac{df}{dr}-\frac{8}{7}\left(  f-1\right)  ^{3}%
\frac{\alpha_3}{r^{3}}-\frac{2}{3}r^{3}\Lambda_{0}-2r\left(  f-1\right)
=0\label{ecudiferentialparaf}%
\end{equation}
It is simpler to work with the function $\varphi\left(  r\right)  $ such that%
\begin{equation}
\varphi\left(  r\right)  =\frac{1-f\left(  r\right)  }{r^{2}}\ .%
\end{equation}
At linear order in the rotation parameters one obtains the equations for
$g_{1}\left(  t,r\right)  $ and $g_{2}\left(  t,r\right)  $ that allow to
construct the slowly rotating black hole solutions. The equations are%
\begin{equation}
A\left(  \varphi,r\right)  \frac{d^{2}g_{i}}{dr^{2}}+B\left(  \varphi
,r\right)  \frac{dg_{i}}{dr}=0\ ,\label{eqsgs}%
\end{equation}
and reduce manifestly to a second order system. Here%
\begin{align}
A(\varphi,r) &  =-\frac{(1-r^{2}\varphi)}{42}\left(  30\kappa(r^{2}%
\varphi^{\prime})^{\prime}+r^{2}(12\kappa\varphi-28\alpha\varphi-7)\right) \ , \\
B(\varphi,r) &  =A^{\prime}(\varphi,r)-\frac{(r^{3}\varphi^{\prime2}%
\varphi+3)}{42r}\left(  30\kappa(r^{2}\varphi^{\prime})^{\prime}%
+r^{2}(12\kappa\varphi-28\alpha\varphi-7)\right)\ .
\end{align}
This simplification is remarkable since a generic perturbation of the static
black hole metric may lead to a fourth-order system \cite{Myers:2010ru},
nevertheless the perturbation that allows constructing the slowly rotating
solutions is determined by a second order system, as in General Relativity and
Lovelock theories.

It is evident that there is a simple solution of the system (\ref{eqsgs})
given by $g_{1}=g_{1}\left(  t\right)  $ and $g_{2}=g_{2}\left(  t\right)  $,
both functions being arbitrary. It is simple to show that at linear order in
the rotation parameters these solutions are pure gauge since one can define%
\begin{equation}
\phi=\tilde{\phi}-\frac{a_{1}}{2}\int g_{1}\left(  t\right)  dt\text{ and
}\psi=\tilde{\psi}-\frac{a_{2}}{2}\int g_{1}\left(  t\right)  dt\ ,
\end{equation}
such that the metric (\ref{slowly}) reduces to a static metric up to terms
which are quadratic on the rotation parameters. This change of coordinates is
obstructed when the functions $g_{1}$ and $g_{2}$ depend on the radial
coordinate. We therefore solve the equations (\ref{eqsgs}) by setting
$g_{1}=g_{1}\left(  r\right)  $ and $g_{2}=g_{2}\left(  r\right)  $. The
equations for the $g_{i}\left(  r\right)  $ can be integrated in terms of
quadratures, since they imply the remarkably compact expression%
\begin{equation}
\frac{g_{i}^{\prime\prime}(r)}{g_{i}^{\prime}(r)}=-\left(  \ln\left[  30\alpha_3
r^{3}(r^{2}\varphi^{\prime})^{\prime}-\frac{28m}{\varphi^{\prime}}\right]
\right)  ^{\prime}\text{ with }i=1,2\ ,
\end{equation}
therefore%
\begin{equation}
g_{i}(r)=B_{i}\int\frac{\varphi^{\prime}dr}{30\alpha_3 r^{3}\varphi^{\prime
}\left(  r^{2}\varphi^{\prime}\right)  ^{\prime}-28m}+C_{i}\label{lasgs}%
\end{equation}
where \thinspace$B_{i}$ and $C_{i}$ are integration constants. It is important
to notice that in absence of the cubic quasitopological term, the expression
for $g_{i}\left(  r\right)  $ can be integrated in a closed form. This was
done in \cite{Kim:2007iw,Yue:2011et}. It is interesting to notice also that the
function $g_{i}\left(  r\right)  $ depends on the Gauss-Bonnet coupling only
through $\varphi$, a feature that is shared by generic Lovelock theories (see
appendix A of \cite{Adair:2020vso}). We have depicted some profiles for $f(r)$ and the off-diagonal functions $g(r)$ in Figure 2.
\begin{figure}[!ht]
     \centering
     \begin{subfigure}[b]{0.45\textwidth}
         \centering
         \includegraphics[width=\textwidth]{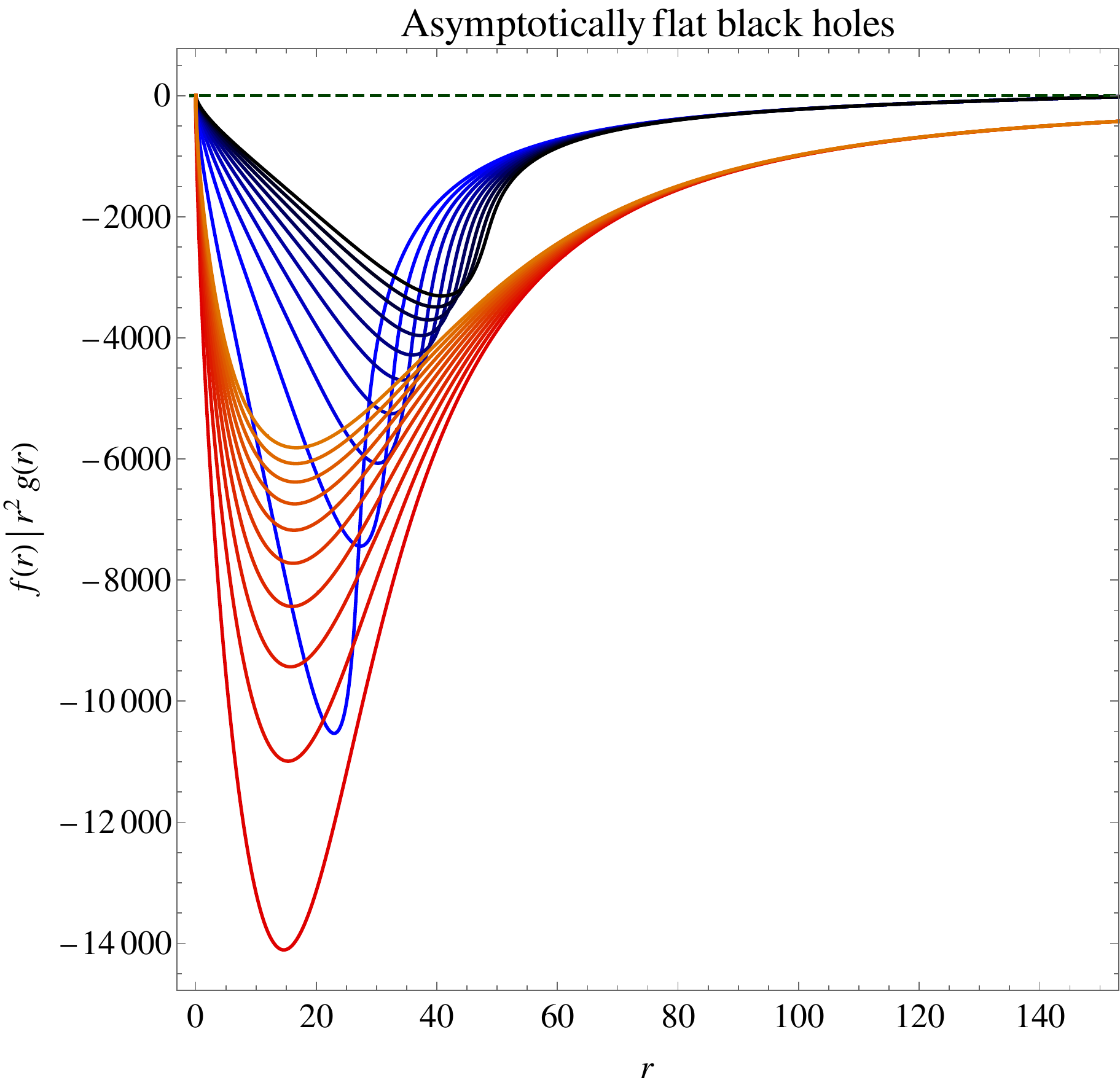}
     \end{subfigure}
     \hfill
     \begin{subfigure}[b]{0.45\textwidth}
         \centering
         \includegraphics[width=\textwidth]{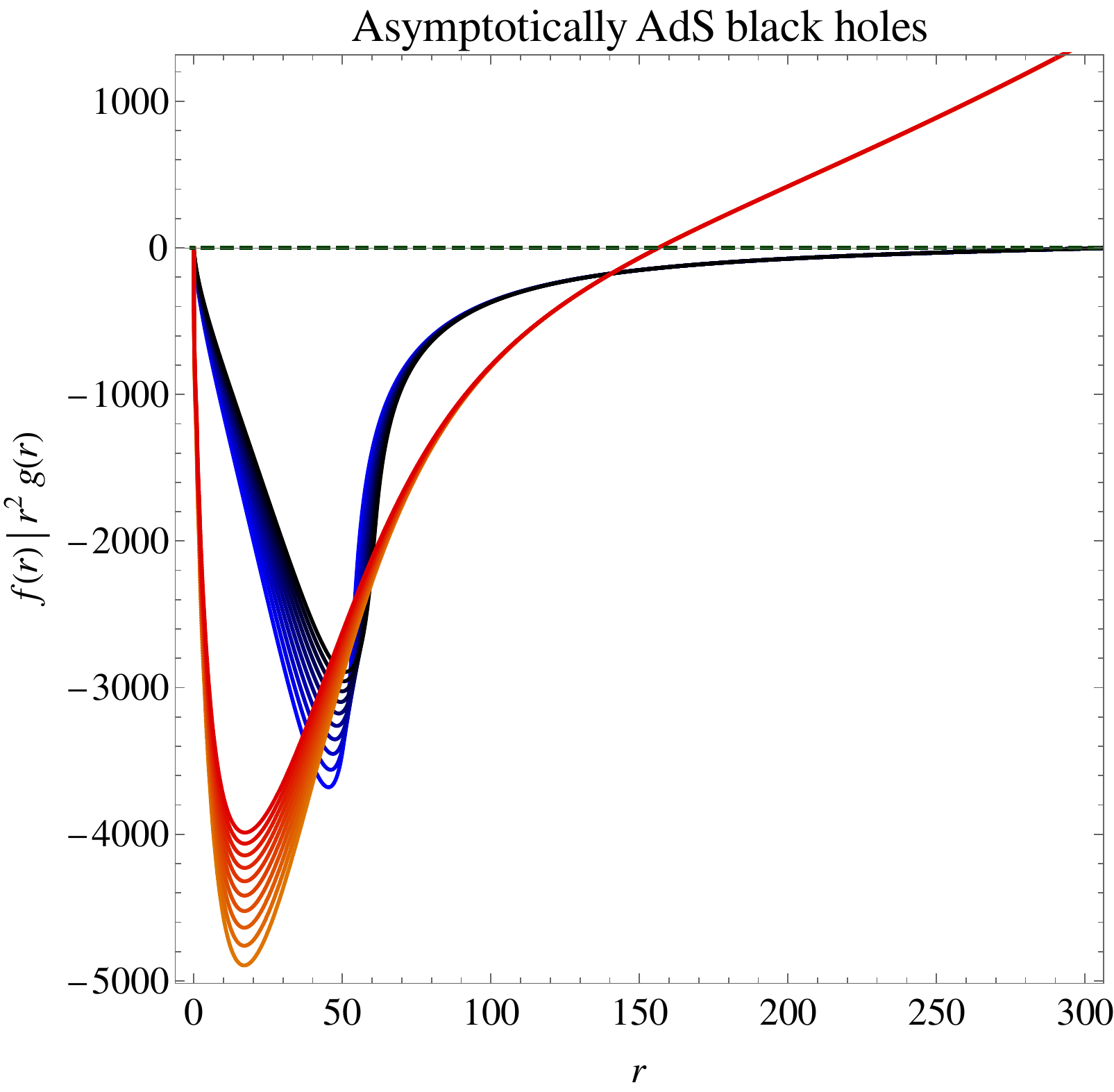}
     \end{subfigure}
    \caption{The figure depicts the profile for $f(r)$ in orange and $g(r)$ in blue for different values of the couplings. We have set the mass parameter as $m=10^7$  and $\alpha_2=\frac{(14+i)}{100}$ while $\alpha_3=-\frac{\alpha_2}{25}$, with $i=1,\dots,10$. For asymptotically AdS black holes the bared cosmological term $\Lambda_{0}=-0.1$}
        \label{AsymBHslowly}
\end{figure}
\section{Slowly rotating solutions from symmetric criticality and quartic
Quasi-topological gravities}

It is well known that the field equations obtained from a reduced Lagrangian,
obtained evaluating $\mathcal{L}=\sqrt{-g}L$ on a particular ansatz, are not always the
correct field equations \cite{Palais:1979rca,Deser:2004gi}. In a spherically symmetric
ansatz, this principle of symmetric criticality (also know as reduced action
or minisuperspace approach), does work correctly in generic diffeomorphism
invariant theories, provided one considers two blackening factors. As pointed
out in \cite{Deser:2004gi}, even in spherical symmetry but without assuming
staticity, in order to obtain a consistent reduced action one must include a
function $h\left(  t,r\right)  $, which is pure gauge%
\begin{equation}
ds^{2}=-f\left(  t,r\right)  dt^{2}+\frac{dr^{2}}{g\left(  t,r\right)
}+h\left(  t,r\right)  dtdr+r^{2}d\Omega^{2}\ .
\end{equation}
The Euler-Lagrange equation associated to $h\left(  t,r\right)  $ in the
Lovelock and cubic, quartic and quintic Quasi-topological gravities is
responsible for the staticity of the solution. Inspired by these facts, one
can show that the four and five-dimensional ansatze%
\begin{align}
ds_{4D}^{2} &  =-N^{2}\left(  r\right)  f\left(  r\right)  dt^{2}+\frac
{dr^{2}}{f\left(  r\right)  }+r^{2}\left(  d\theta^{2}+\sin^{2}\theta
d\phi^{2}\right)  +h_{1}\left(  r,\theta\right)  dtd\phi\ ,\\
ds_{5D}^{2} &  =-N^{2}\left(  r\right)  f\left(  r\right)  dt^{2}+\frac
{dr^{2}}{f\left(  r\right)  }+r^{2}\left(  d\theta^{2}+\sin^{2}\theta
d\phi^{2}+\cos^{2}\theta d\psi^{2}\right)  +\nonumber\\
&  g_{1}\left(  r,\theta\right)  dtd\phi+g_{2}\left(  r,\theta\right)
dtd\psi\ ,\label{5dfin}%
\end{align}
do indeed lead to the correct field equations in the slowly rotating case for
General Relativity, Quadratic Gravity, as well as for a generic cubic
combination (even beyond the Quasi-topological Lagrangians). Notice that it is
important not to fix the $\theta$ dependence \textit{a priori} on the $g_{i}$. With
this evidence at hand, as a matter of fact we apply the principle of symmetric
criticality for the quartic Quasi-topological theories in the slowly rotating
regime along the ansatz (\ref{5dfin}). It is known that on spherically
symmetric spacetimes there is a degeneracy on quartic Quasi-topological
theories, which as shown in the next section is partially removed by requiring
second order field equations for the slowly rotating ansatz, leading to a
quartic Quasi-topological theory with special properties, even beyond
spherical symmetry and considering fluctuations around the spherically
symmetric black hole solutions.

\section{Quartic quasitopological theory}

In general for dimension $d\geq8$, there are $26$ independent scalar terms of
the form $Riem^{4}$. Due to multiple curvature identities that appear as one
lowers the dimension, the number of independent $Riem^{4}$ combinations is
reduced \cite{Fulling:1992vm}, and in five dimensions such list reduces to $20$
independent combinations. In order to be able to establish relations in the
future in a simple manner between quartic combinations with special properties
in different dimensions, even though we will focus on the five dimensional
case, we will work with the complete, redundant, list of $26$ scalars. The
specific basis of quartic curvature terms $L_{I}$ is given in
Appendix A. The action containing up to the quartic Quasi-topological theory
has the form%
\begin{align}
I  & =I_{CQTG}+I_{QQTG}^{\left(  4\right)  }\nonumber\\
& =I_{CQTG}+\int\sqrt{-g}d^{5}x\sum_{I=1}^{26}d_{I}L_{I}%
\label{actionquartic}%
\end{align}
where we have defined $I_{CQTG}$ in equation (\ref{LaCubica}).

The quartic Quasi-topological theories are defined such that the field
equations on spherical symmetry reduce to a first order equation for the
function $f\left(  r\right)$, as it occurs in Lovelock theories. This
imposes $11$ independent contraints on the relative couplings $d_I$ of the quartic
terms. One can further reduce the number of independent couplings by requiring
the absence of ghosts when the theory is linearized around a maximally
symmetric AdS vacuum. Following the linearization procedure developed in
reference \cite{Bueno:2016ypa}, one can see that for generic higher curvature
combinations, the linearized equations around AdS contain two higher
derivative contributions of the form%
\begin{align}
    \left(  2a+c\right)  \bar{\square}G_{\mu\nu}^{L}+\left(  a+2b+c\right)
\left(  \bar{g}_{\mu\nu}\bar{\square}-\bar{\nabla}_{\mu}\bar{\nabla}_{\nu
}\right)  R^{L}+...=0 ,\label{lineareqn}
\end{align}
where the parameters $a$, $b$ and $c$ depend on the specific Lagrangian (see
\cite{Bueno:2016ypa}) and ($...$) stands for second order terms. Requiring the absence of
ghosts implies $2$ further independent linear restrictions on the couplings.
In summary, the $26$ couplings $d_{\left(  i\right)  }$ are restricted by $13$
independent constraints, leading to $13$ remaining couplings. Notice that, as mentioned above, some redundancy coming
from algebraic identities of the Riemann tensor has not been removed.
Remarkably, after imposing all these constraints, one still has a non-trivial theory.

Now we provide a new argument that allows to further restrict these couplings.
Using the reduced action principle described above by the metric ansatz
(\ref{5dfin}), on the theory constrained by the $13$ conditions coming from
spherically symmetry and absence of ghosts around AdS, one obtains an equation
for the off-diagonal metric components $g_{1}=\mu^{2}h_{1}\left(  r\right)  $
and $g_{2}=\left(  1-\mu^{2}\right)  h_{2}\left(  r\right)  $ of the form%
\begin{equation}
\xi_{1}\frac{d^{4}h_{i}}{dr^{4}}+\xi_{2}\frac{d^{3}h_{i}}{dr^{3}}+\xi_{3}%
\frac{d^{2}h_{i}}{dr^{2}}+\xi_{4}\frac{dh_{i}}{dr}=0\ ,\text{ }i=1,2
\end{equation}
where the $\xi_{j}$ are functions of $r$, $f\left(  r\right)  $ and the
couplings. Requiring the absence of the fourth and third order derivatives
implies a single new constraint on the couplings, leading to a non-trivial
theory with special properties even beyond spherical symmetry.

In summary, requiring the existence of simple spherically symmetric black
holes with a single metric function governed by a first order equation, plus
the absence of ghosts around AdS, in addition to the existence of slowly
rotating black holes with an off-diagonal metric function determined by a
second order equation, leads to $14$ relations between the couplings (see
Appendix \ref{B}).

\bigskip

When the rotation is turned-off, as shown in \cite{Dehghani:2011vu}, the equation for
the blackening factor of the spherically symmetric black holes turns out to be
restricted by the following polynomial equation%
\begin{equation}
21\Bar{\alpha}_{4}\left( f-1\right)  ^{4}-24\alpha_{3}(f-1)^{3}r^{2}%
-84\alpha_{2}r^{4}(f-1)^{2}+42(f-1)r^{6}+7\Lambda_{0}r^{8}%
=14mr^{4}\label{polyf2}%
\end{equation}
Where $\Bar{\alpha}_4$ is a combination of the cuartic invariants given below in equation (\ref{a4b}) As before, for generic values of the couplings of the terms with different
order in the curvature, the quartic theory admits four different vacua with
four different values of the effective cosmological constant $\Lambda
_{\text{eff}}^{\left(  i\right)  }$ with $i=1,2,3,4$. The four solutions of
(\ref{polyf2}) will asymptotically behave as%
\begin{equation}
f_{\left(  i\right)  }\left(  r\right)  =-\frac{\Lambda_{\text{eff}}^{\left(
i\right)  }}{6}r^{2}+1-\frac{m}{r^{2}}+...\ .
\end{equation}
Equation (\ref{polyf2}) comes from the integration of the following
differential equation%
\begin{align}
\frac{\delta\mathcal{L}}{\delta N} &  =\left(  -\frac{2\Bar{\alpha}%
_{4}(f-1)^{3}}{r^{4}}+\frac{12(f-1)^{2}\alpha_{3}}{7r^{2}}+4(f-1)\alpha
_{2}-r^{2}\right)  f^{\prime}\nonumber\\
&  +\frac{2\Bar{\alpha}_{4}(f-1)^{4}}{r^{5}}-\frac{8\alpha_{3}(f-1)^{3}%
}{7r^{3}}-2r(f-1)-\frac{2}{3}\Lambda_{0}r^{3}%
\end{align}
which as indicated comes from the variation of the reduced action principle,
with respect to the function $N$ in the metric (\ref{5dfin}).

Remarkably, in the context of the quartic Quasi-topological theories one can
further relate the couplings such that the equation for the off-diagonal
components of the metric $g_{\left(  i\right)  }\left(  r\right)  $ take a
very simple, second order form, leading to%
\begin{align}
\frac{g_{i}^{\prime\prime}}{g_{i}^{\prime}} &  =-\left(  \frac{3}{r}\right.
\nonumber\\
&  \left.  +\frac{\left(  7r^{2}(r^{2}\beta_{1}\varphi^{\prime\prime2}%
+r\beta_{2}\varphi^{\prime}\varphi^{\prime\prime}+\beta_{3}\varphi^{\prime
2})+(70\Bar{\alpha}_{4}\varphi+60\alpha_{3})(r^{2}\varphi^{\prime})^{\prime
}-\frac{28m}{r^{3}\varphi^{\prime}}\right)  ^{\prime}}{7r^{2}(r^{2}\beta
_{1}\varphi^{\prime\prime2}+r\beta_{2}\varphi^{\prime}\varphi^{\prime\prime
}+\beta_{3}\varphi^{\prime2})+(70\Bar{\alpha}_{4}\varphi+60\alpha_{3}%
)(r^{2}\varphi^{\prime})^{\prime}-\frac{28m}{r^{3}\varphi^{\prime}}}\right)
\\
&  =-\left(  \ln\left[  7r^{5}(r^{2}\beta_{1}\varphi^{\prime\prime2}%
+r\beta_{2}\varphi^{\prime}\varphi^{\prime\prime}+\beta_{3}\varphi^{\prime
2})+r^{3}(70\Bar{\alpha}_{4}\varphi+60\alpha_{3})(r^{2}\varphi^{\prime
})^{\prime}-\frac{28m}{\varphi^{\prime}}\right]  \right)  ^{\prime}\\
g_{i} &  =B_{i}\int\frac{\varphi^{\prime}dr}{7r^{5}\varphi^{\prime}\left(
\beta_{1}r^{2}\varphi^{\prime\prime2}+r\beta_{2}\varphi^{\prime}%
\varphi^{\prime\prime}+\beta_{3}\varphi^{\prime2}\right)  +r^{3}%
\varphi^{\prime}(70\Bar{\alpha}_{4}\varphi+60\alpha_{3})(r^{2}\varphi^{\prime
})^{\prime}-28m}+C_{i}\label{gdelcuartico}%
\end{align}
where
\begin{align}
\Bar{\alpha}_{4} &  =\alpha_{4}\left(  d_{13}+2d_{16}+4d_{17}+8d_{18}%
+48d_{19}+d_{20}+4d_{21}+8d_{22}+6d_{23}\right.\nonumber\\
&\quad\quad\left.+22d_{24}+44d_{25}+128d_{26}\right)\\
\beta_{1} &  =\alpha_{4}\left(  14d_{20}-30d_{21}+168d_{22}+30d_{23}%
+303d_{24}-90d_{25}-3504d_{26}\right.\nonumber\\
&\quad\quad\left.+18d_{17}+98d_{18}+216d_{19}+30d_{13} +44d_{16}-16d_{4}\right) \label{a4b} \\
\beta_{2} &  =\alpha_{4}\left(  156d_{13}+248d_{16}+128d_{17}+604d_{18}%
+1632d_{19}+83d_{20}-88d_{21}\right.\nonumber\\
&\quad\quad\left.+956d_{22}+208d_{23}+1828d_{24}+48d_{25}-16960d_{26}-64d_{4}\right)  \\
\beta_{3} &  =\alpha_{4}\left(  231d_{13}+398d_{16}+268d_{17}+1034d_{18}
+3232d_{19}+138d_{20}-38d_{21}\right.\nonumber\\
&\quad\quad\left.+1566d_{22}+428d_{23}+3123d_{24}+908d_{25} -23040d_{26}-64d_{4}\right)
\end{align}

Equation (\ref{gdelcuartico}) is remarkably simple. The quartic contributions
to the denominator of the integrand are of four types. First, there is an
$\bar{\alpha}_{4}$ contribution, which adds a term to the previous cubic
Quasi-topological contribution. Notice that $\bar{\alpha}_{4}$ controls the
contribution of quartic Quasi-topological to the spherically symmetric
solution (\ref{polyf2}). It is tempting to conjecture a pattern here for
Quasi-topological gravities of higher order. On top of this, the quartic
Quasi-topological family add three new, independent terms of different kind to
the denominator of the integrand of (\ref{gdelcuartico}). We have checked the
linear independence of these combinations after using the $14$ constraints
mentioned above. One may set zero each of the $\beta s$, further restricting
the couplings in favor of the simplicity of the slowly rotating solution. This would lead to the three-parameter family of theories mentioned in the abstract.

\bigskip

In the next sections we move beyond the slowly rotating approximations for the
cubic Quasi-topological theories, within the Kerr-Schild Ansatz.

\section{Beyond slowly rotating solutions: Kerr-Schild Ansatz}

The Kerr-Schild (KS) ansatz has been a very fruitful arena for the
construction of rotating black holes in General Relativity in arbitrary
dimensions. The whole family of rotating black holes with horizons of
hyperspherical topology can be casts in this form \cite{Hawking:1998kw,Myers:1986un,Carter:1968ks,Gibbons:2004js,Gibbons:2004uw,Ayon-Beato:2015nvz}, both in the asymptotically flat case as
well as in the presence of a cosmological constant. The Kerr-Schild ansatz is
defined by
\begin{equation}
g_{ab}=\tilde{g}_{ab}+F\left(  x\right)  k_{a}k_{b}\label{KS}%
\end{equation}
where the metric $\tilde{g}_{ab}$ is a seed metric, while the vector field $k$
defines a null and geodesic congruence on the seed spacetime. The latter
properties lead to an Einstein tensor that is linear in the function $F$ and
its derivatives, facilitating the integration of the field equations.
Depending on the expansion, twist and shear of the congruence, the KS ansatz
can accommodate rotating or static black holes as well as pp-waves or (A)dS
waves. This ansatz has even allowed to construct an intrinsically rotating
solution in five dimensions in the presence of a term that is quadratic in the
curvature in the context of the Einstein-Gauss-Bonnet theory with a single
(A)dS solution \cite{Anabalon:2009kq}, and has been explored in some generality in
Lovelock gravities in \cite{Ett:2011fy}\footnote{See also
\cite{Babichev:2020qpr} for the constructions of black holes
using the KS ansatz in DHOST theories, as well as \cite{Cisterna:2015uya}
for the construction of slowly rotating black holes in Horndeski theories.}.
Since the Kerr-Schild ansatz has been useful to construct rotating solutions
in higher curvature theories, here we explore this metric ansatz in the
context of the cubic Quasi-topological gravity.

The Kerr-Schild ansatz leading to rotating solutions in General Relativity in
five dimensions has a seed%
\begin{align}
d\Tilde{s}^{2}= &  -\frac{\left(  1+\frac{r^{2}}{\ell^{2}}\right)
\Delta_{\theta}dt^{2}}{\Xi_{a}\Xi_{b}}+\frac{r^{2}\rho^{2}dr^{2}}{\left(
1+\frac{r^{2}}{\ell^{2}}\right)  (r^{2}+a^{2})(r^{2}+b^{2})}+\frac{\rho
^{2}d\theta^{2}}{\Delta_{\theta}}\nonumber\\
&  +\frac{r^{2}+a^{2}}{\Xi_{a}}\sin^{2}\theta d\phi^{2}+\frac{r^{2}+b^{2}}%
{\Xi_{b}}\cos^{2}\theta d\psi^{2}%
\end{align}
with $\Xi_{a}=1-\frac{a^{2}}{\ell^{2}}$, $\Xi_{b}=1-\frac{b^{2}}{\ell^{2}}$,
$\rho^{2}=r^{2}+a^{2}\cos^{2}\theta+b^{2}\sin^{2}\theta$, $\Delta_{\theta}%
=\Xi_{a}\cos^{2}\theta+\Xi_{b}\sin^{2}\theta$ and null and geodesic vector%
\begin{equation}
k_{a}dx^{a}=\frac{\Delta_{\theta}dt}{\Xi_{a}\Xi_{b}}+\frac{r^{2}\rho^{2}%
dr}{\left(  1+\frac{r^{2}}{\ell^{2}}\right)  (r^{2}+a^{2})(r^{2}+b^{2})}%
-\frac{a\sin^{2}\theta d\phi}{\Xi_{a}}-\frac{a\cos^{2}\theta d\psi}{\Xi_{b}}%
\end{equation}
Here the seed metric is that of an AdS spacetime of curvature radius $\ell$,
while $a$ and $b$ are the rotation parameters. Since we will be interested in
rotating solutions for arbitrary values of the couplings, the leading
asymptotic behavior of the metric in Quasi-topological gravity should be that
of GR, with one of the effective cosmological constants. In GR the function
$F\left(  r,\theta\right)  $ is given by%
\begin{equation}
F\left(  r,\theta\right)  =F_{GR}\left(  r,\theta\right)  =\frac{2M}%
{r^{2}+a^{2}\cos^{2}\theta+b^{2}\sin^{2}\theta}.
\end{equation}
For simplicity hereafter we focus on the case with $a=b$. This choice enlarges
the isometry group of the solution from $\mathbb{R}
_{t}\times U\left(  1\right)  ^{2}$ to $\mathbb{R}
_{t}\times SU\left(  2\right)  $. In this case, asymptotically one has%
\begin{equation}
F_{GR}\left(  r,\theta\right)  =F_{GR}\left(  r\right)  =\frac{2M}{r^{2}%
}-\frac{Ma^{2}}{r^{4}}+O\left(  r^{-6}\right)  \ .
\end{equation}
As it occurs for spherically symmetric black holes, one requires the
asymptotic behavior of the function $F\left(  r\right)  $ to match that of GR
with an effective cosmological constant, implying that in the presence of
generic higher curvature terms, one must assume an expansion of the form%
\begin{equation}
F\left(  r\right)  =r^{-2}\left(  A_{2}+\frac{A_{3}}{r}+\frac{A_{4}}{r^{2}%
}+\frac{A_{5}}{r^{3}}+O\left(  r^{-5}\right)  \right)  \ ,\label{expanF}%
\end{equation}
at infinity.

In cubic Quasi-topological gravity the trace of the field equations is a
second order constraint on the metric, for arbitrary spacetimes. In the case
of the Kerr-Schild ansatz (\ref{KS}), with $a=b$ and therefore assuming
$F=F\left(  r\right)  $, such constraint reduces to%
\begin{align}
0 &  =-30l^{2}(r^{2}+a^{2})^{2}\left[  \frac{l^{4}a^{2}}{30}\alpha_3\left(
a^{2}-5r^{2}\right)  \left(  a^{2}+r^{2}\right)  ^{2}F^{\prime2}+\frac
{2l^{4}a^{2}r\alpha_3}{3}\left(  r^{4}-1\right)  FF^{\prime}\right.  \nonumber\\
&  +r^{2}\left(  \frac{8\alpha_3 l^{4}}{15}\left(  a^{2}-\frac{3}{4}%
r^{2}\right)  \left(  a^{2}-r^{2}\right)  F^{2}-\frac{4l^{2}}{3}\left(
l^{2}\alpha_2-\frac{\alpha_3}{5}\right)  \left(  a^{2}+r^{2}\right)  ^{2}\left(
a^{2}-3r^{2}\right)  F\right.  \nonumber\\
&  \left.  +\left(  a^{2}+r^{2}\right)  ^{4}\left(  l^{4}-4l^{2}\alpha_2
+\frac{2}{5}\alpha_3\right)  \right]  F^{\prime\prime}+14\alpha_3\left(
a^{2}-\frac{5r^{2}}{7}\right)  rl^{6}\left(  a^{2}+r^{2}\right)  ^{3}%
a^{2}F^{\prime3}\nonumber\\
&  +40l^{4}\left(  a^{2}+r^{2}\right)  ^{2}\left[  \frac{\alpha_3 l^{2}}%
{4}\left(  a^{6}-\frac{46}{5}a^{4}r^{2}+\frac{33}{5}a^{2}r^{4}-\frac{12}%
{5}r^{6}\right)  F\right.\nonumber\\
&\left.+r^{2}l^{2}\left(  \alpha_2-\frac{\alpha_3}{5l^{2}}\right)  \left(
a^{2}+r^{2}\right)  ^{2}\left(  a^{2}-3r^{2}\right)  \right]  F^{\prime
2}\nonumber\\
&  -60rl^{2}\left(  a^{2}+r^{2}\right)  ^{5}\left[  \left(  a^{2}%
+3r^{2}\right)  \left(  l^{4}-4l^{2}\alpha_2+\frac{2\alpha_3}{5}\right)
+\frac{8\alpha_3 l^{4}\left(  a^{2}-3r^{2}\right)  }{15\left(  a^{2}%
+r^{2}\right)  ^{4}}\left(  a^{4}-\frac{5}{2}a^{2}r^{2}+\frac{r^{4}}%
{4}\right)  F^{2}\right.  \nonumber\\
&  \left.  -\frac{4l^{2}\left(  l^{2}\alpha_2-\frac{\alpha_3}{5}\right)
}{3\left(  a^{2}+r^{2}\right)  ^{2}}\left(  a^{4}-10a^{2}r^{2}-3r^{4}\right)
F\right]  F^{\prime}+20r^{2}\left(  a^{2}+r^{2}\right)  ^{6}\bigg[  \left(
5l^{6}\Lambda_{0}+30l^{4}-60l^{2}\alpha_2+4\alpha_3\right)   \nonumber\\
&   -\frac{2\alpha_3 r^{2}l^{6}\left(  25a^{4}-22a^{2}r^{2}+r^{4}\right)
}{5\left(  a^{2}+r^{2}\right)  ^{6}}F^{3}-\frac{24l^{4}(5l^{2}\alpha_2
-\alpha_3)(3a^{2}-r^{2})a^{2}}{15(a^{2}+r^{2})^{4}}F^{2}\nonumber\\
&-\frac{9l^{2}\left(
5l^{4}-20l^{2}\alpha_2+2\alpha_3\right)  }{5(a^{2}+r^{2})}F\bigg]  \label{traza}%
\end{align}
The equation $E_{\ \theta}^{\theta}=0$ coming from the action
(\ref{actionquartic}) is of fourth order, and it is useful in the following
analysis, but since it is not very illuminating in its complete form we do not
provide it here. With these two equations at hand we can move forward.
Assuming an expansion of the form (\ref{expanF}) in the $E_{\ \theta}^{\theta
}=0$ equation, the leading and subleading orders in the $r\rightarrow\infty$
expansion lead to%
\begin{equation}
\Upsilon\left[  \Lambda\right]  =\Lambda^{3}\alpha_3+90\Lambda^{2}%
\alpha_2+270\Lambda-270\Lambda_{0}=0\ ,\ \frac{d\Upsilon\left[  \Lambda\right]
}{d\Lambda}A_{3}=0\ .
\end{equation}
Requiring the genericity of the couplings imply $\frac{d\Upsilon\left[
\Lambda\right]  }{d\Lambda}$ $\neq0$, namely the curvature radius of the seed
is a simple zero of the polynomial $\Upsilon\left[  \Lambda\right]  $.
Consequently $A_{3}=0$. Considering this information in the trace
(\ref{traza}) further leads to $A_{5}=A_{7}=0$ as well as  $A_{2n+1}=0$. One also obtains as a consequence of both equations that%
\begin{equation}
\Upsilon^{\prime}\left[  \Lambda\right]  \left(  a^{2}A_{2}+A_{4}\right)
=0\ .
\end{equation}
Here we must emphasize that if one imposes relations between the couplings
such that $\Upsilon^{\prime}\left[  \Lambda\right]  =0$, new branches of
solutions may appear. Actually, from what is known from the spherically
symmetric black holes in Lovelock theories with a unique vacuum, it would be
natural to expect an asymptotic behavior of the form $r^{-\alpha}$ with
$\alpha$ a non-integer number \cite{Crisostomo:2000bb}.

To the next order in the $r\rightarrow\infty$ expansion one gets%
\begin{equation}
\Upsilon^{\prime}\left[  \Lambda\right]  A_{6}=\left(  a^{4}\Upsilon^{\prime
}\left[  \Lambda\right]  -3\Upsilon^{\prime\prime}\left[  \Lambda\right]
\right)  A_{2}\ .
\end{equation}
Finally, the following subleading order in the expansion as $r\rightarrow
\infty$, leads to an inconsistency, since it implies%
\begin{align}
\Upsilon^{\prime}\left[  \Lambda\right]  A_{8} &  =-a^{6}\Upsilon^{\prime}\left[  \Lambda\right]
A_{2}+13a^{2}\Upsilon^{\prime\prime}\left[  \Lambda\right]  A_{2}^{2}\\
&  =-a^{6}\Upsilon^{\prime
}\left[  \Lambda\right]  A_{2}+\frac{A_{2}^{2}}{7}(7644\Lambda\alpha_3
-8280\alpha_2+109\Upsilon^{\prime\prime}\left[  \Lambda\right]  )a^{2}\ .
\end{align}
If one want $A_{2}$ to be arbitrary, since this might be later identified with
the mass of the rotating black hole, the latter equation implies new
constraints that relate the couplings of different orders in the curvature, which
is incompatible with our previous assumption of genericity of the couplings.
Consequently, we have proved that it is impossible to accommodate an
asymptotically Kerr-AdS$_{5}$, rotating solution of cubic Quasi-topological
theory within the Kerr-Schild ansatz, for generic values of the couplings when
the two rotation parameters are equal.

\bigskip

\section{Conclusions}

In this paper we have constructed the slowly rotating solutions of cubic and
quartic Quasi-topological gravities. The former is unique, while the latter is
not. Namely, restricting first order field equations on spherical symmetry
singles out a unique cubic theory, which is also ghost-free around AdS, while
in the quartic theory both constraints lead to 13 restrictions on the 20
algebraically independent, $Riem^{4}$ scalars in five dimensions. We have show
that the slowly rotating solution in the cubic theory is governed by a second
order equation, while in quartic Quasi-topological theories, requiring the
equations to be of second order for the slowly rotating metric, implies one
extra constraints on the couplings, leading to 14 conditions and therefore to
a 6-parameter family of Lagrangians. Even more, from the explicitly
integration of the off-diagonal components we have seen that three extra
constraints could be imposed in order to achieve a simple off-diagonal metric
functions, which allows an easy conjectural form for Quasi-topological
gravities of arbitrary high order. In these computations we have worked with
the redundant list of 26 quartic scalars (the family $\mathcal{R}_{8,4}^{0}$
in the notation of \cite{Fulling:1992vm}), since in this manner it will be now
easy to find relations between Lagrangians of quartic theories in different
dimensions. The same strategy allowed to prove that the dimensional reduction
of the single cubic Quasi-topological theory in five dimensions
\cite{Cisterna:2018tgx}, leads to the cubic combination with second order
equations on homogeneous and isotropic cosmologies in dimensions four
identified and explored in \cite{Arciniega:2018fxj,Arciniega:2018tnn,Arciniega:2019oxa,Edelstein:2020nhg}. It
was also proved in \cite{Cisterna:2018tgx} that the scalar gravitational
perturbations on these FLRW background are governed by equations that are of
second order in time. More recent explorations have shown that for vector and
tensor modes this property might be absent in the cubic theory \cite{Pookkillath:2020iqq}. It would be interesting to explore whether these
connections extend to the context of quartic theories, and beyond. The freedom
that still remains in the couplings can be used in ones favour at the moment
of selecting gravitational theories with sensible properties. It is also
interesting to mention that the classification of duality invariant
$\alpha^{\prime}$ corrections in string inspired scenarios, in the
cosmological ansatz, has lead to combinations of higher curvature terms with
similar properties, namely leading to second order equations for the FLRW
ansatz \cite{Hohm:2019jgu}. Also in the context of
$\alpha^{\prime}$ corrections of String Theory it was proved in
\cite{Bueno:2019ltp} that using the freedom of field redefinitions,
intrinsic to the perturbative approach, every higher curvature combination can
be rewritten in the frame of Einsteinian gravities, which also possess special
properties on spherical symmetries \cite{Bueno:2016xff,Hennigar:2016gkm,Bueno:2016lrh,Bueno:2017sui,Bueno:2018xqc,Cano:2019ozf,Adair:2020vso,Hennigar:2017ego}. In summary, there seem to be a network of theories with
special properties in different dimensions, which deserves to be
explored in its own right, and even more in the context of higher curvature,
perturbative corrections to GR.

\section{Acknowledgments}

We thank Pablo Bueno, Jos\'{e} Edelstein and Nicol\'{a}s Grandi for
enlightening comments. We specially thank Adolfo Cisterna for suggesting to
analyze the equations of Quasi-topological gravities in the slowly rotating
regime. This work was partially funded by FONDECYT grant 1181047 and FONDECYT
grant 11191175. O.F. would like to thank to the Direcci\'{o}n de Investigación and Vicerrector\'{\i}a de Investigaci\'{o}n of the Universidad Cat\'{o}lica de la Sant\'{\i}sima Concepci\'{o}n, Chile, for their constant support in particular through the Project DINREG 19/2018 of the Direcci\'{o}n de Investigaci\'{o}n of the Universidad Cat\'{o}lica de la Sant\'{\i}sima Concepci\'{o}n, Chile. N.M. thanks the support of Vicerrector\'{\i}a de Investigaci\'{o}n, UdeC.

\bigskip

\appendix
\section{\label{A}}

The following list defined the $26$ invariants which are independent in
dimension greater or equal than eight. These correspond to the family $\mathcal{R}_{8,4}^{0}$
in the notation of \cite{Fulling:1992vm}:%

\begin{align}
&  L_{1}=R^{pqbs}R_{p\ b}^{\ a\ u}R_{a\ q}^{\ v\ w}R_{uvsw},\ L_{2}%
=R^{pqbs}R_{p\ b}^{\ a\ u}R_{a\ u}^{\ v\ w}R_{qvsw},\ L_{3}=R^{pqbs}%
R_{pq}^{\ \ au}R_{b\ a}^{\ v\ w}R_{svuw},\nonumber\\
&  L_{4}=R^{pqbs}R_{pq}^{\ \ au}R_{ba}^{\ \ vw}R_{suvw},\ L_{5}=R^{pqbs}%
R_{pq}^{\ \ au}R_{au}^{\ \ vw}R_{bsvw},\ L_{6}=R^{pqbs}R_{pqb}^{\ \ \ a}%
R_{\ \ \ s}^{uvw}R_{uvwa},\nonumber\\
&  L_{7}=\left(  R^{pqbs}R_{pqbs}\right)  ^{2},\ L_{8}=R^{pq}R^{bsau}%
R_{b\ ap}^{\ v}R_{svuq},\ L_{9}=R^{pq}R^{bsau}R_{bs\ p}^{\ \ v}R_{auvq}%
,\nonumber\\
&  L_{10}=R^{pq}R_{p\ q}^{\ b\ s}R_{\ \ \ b}^{auv}R_{auvs},\ L_{11}%
=RR^{pqbs}R_{p\ b}^{\ a\ \ u}R_{qasu},\ L_{12}=RR^{pqbs}R_{pq}^{\ \ au}%
R_{bsau},\nonumber\\
&  L_{13}=R^{pq}R^{bs}R_{\ p\ b}^{a\ u}R_{aqus},\ L_{14}=R^{pq}R^{bs}%
R_{\ p\ q}^{a\ u}R_{abus},\ L_{15}=R^{pq}R^{bs}R_{\ \ pb}^{au}R_{auqs}%
,\nonumber\\
&  L_{16}=R^{pq}R_{p}^{\ b}R_{\ \ \ q}^{sau}R_{saub},\ L_{17}=R^{pq}%
R_{pq}R^{bsau}R_{bsau},\ L_{18}=RR^{pq}R_{\ \ \ p}^{bsa}R_{bsaq},\nonumber\\
&  L_{19}=R^{2}R^{pqbs}R_{pqbs},\ L_{20}=R^{pq}R^{bs}R_{b}^{\ a}%
R_{psqa},\ L_{21}=RR^{pq}R^{bs}R_{pbqs},\nonumber\\
&  L_{22}=R^{pq}R_{p}^{\ b}R_{q}^{\ s}R_{bs},\ L_{23}=\left(  R^{pq}%
R_{pq}\right)  ^{2},\ L_{24}=RR^{pq}R_{p}^{\ b}R_{qb},\ L_{25}=R^{2}%
R^{pq}R_{pq},\ L_{26}=R^{4}. \label{quarticinvariants}%
\end{align}

\section{\label{B}}

On spherical symmetry, there are eleven independent constraints on the couplings of the quartic combinations, that lead to a first order equation for the function $f(r)$. The constraints are:
\begin{align}
    0 &= 4d_4+4d_6+d_{20}+2d_{15}+4d_{18}+2d_{23}+2d_{10}+8d_{19}+2d_{16}+4d_9+4d_{25}+2d_2\nonumber\\
&\quad+8d_5+8d_7+2d_{21}+2d_{13}+d_{14}+4d_{17}+8d_{26}+d_{22}+d_1+2d_{24}+8d_{12}\\
0 &= 2d_{10}+16d_{12}+4d_{13}+2d_{14}+4d_{15}+4d_{16}+8d_{17}+12d_{18}+32d_{19}+3d_{20}+8d_{21}\nonumber\\
&\quad+4d_{22}+8d_{23}+10d_{24}+24d_{25}+64d_{26}+4d_9\\
0 &= 4d_{12}+2d_{18}+8d_{19}+d_{21}+d_{24}+4d_{25}+16d_{26}\\
0 &= 4d_2+28d_{18}+54d_{24}+38d_{21}+2d_3+32d_7+576d_{26}+112d_{19}+7d_{13}+9d_{14}+4d_1+8d_6\nonumber\\
&\quad+164d_{25}+2d_8+6d_{11}+18d_{22}+28d_{17}+44d_{23}+8d_{16}+11d_{20}+6d_{10}+6d_{15}
\end{align}
\begin{align}
0 &= 2d_{10}+2d_{14}+8d_{17}+6d_{18}+48d_{19}+d_{20}+10d_{21}+8d_{23}+9d_{24}+56d_{25}+288d_{26}\\
0 &= 3d_{17}+8d_{19}+2d_{23}+5d_{25}+36d_{26}+4d_{7}\\
0 &= 25d_1+4d_{10}-36d_{11}+47d_{13}+32d_{15}+62d_{16}-148d_{17}+66d_{18}-768d_{19}+30d_2\nonumber\\
&\quad+30d_{20}-78d_{21}+135d_{22}-82d_{23}+123d_{24}-916d_{25}-7488d_{26}+8d_3+68d_6\\
0 &= d_1+12d_{11}+11d_{13}+8d_{15}+14d_{16}+12d_{17}+62d_{18}+288d_{19}-2d_2+12d_{20}+54d_{21}\nonumber\\
&\quad+39d_{22}+30d_{23}+117d_{24}+300d_{25}+1152d_{26}+4d_6\\
0&= 6d_1-14d_{10}+24d_{13}+12d_{15}+36d_{16}-76d_{17}+22d_{18}-384d_{19}+12d_2+21d_{20}\nonumber\\
&\quad+6d_{21}+102d_{22}+44d_{23}+153d_{24}-4d_{25}-960d_{26}+24d_6\\
0&= 14d_{10}+18d_{13}-12d_{15}+48d_{16}+148d_{17}+134d_{18}+768d_{19}-24d_2+15d_{20}+30d_{21}\nonumber\\
&\quad+180d_{22}+136d_{23}+393d_{24}+556d_{25}+576d_{26}\\
0&= d_{10}+2d_{17}+4d_{18}+24d_{19}-4d_{23}+3d_{24}-10d_{25}-144d_{26}.
\end{align}
From (\ref{lineareqn}) there are two new constraints such that the linearized equations around AdS are second order, these are
\begin{align}
    0 &= -8d_{13}-6d_{15}-10d_{16}-22d_{17}-46d_{18}-248d_{19}-7d_{20}-16d_{21}-28d_{22}-16d_{23}\nonumber\\
    &\quad-71d_{24}-74d_{25}+272d_{26}\\
    0&=  -4d_{13}-4d_{15}-4d_{16}-8d_{17}-22d_{18}-112d_{19}-5d_{20}-14d_{21}-12d_{22}-12d_{23}\nonumber\\
    &\quad-39d_{24}-72d_{25}+96d_{26}.
\end{align}
The new constraint, that leads to a second order equation for the off-diagonal metric component of the slowly rotating metric  $h_i$, is
\begin{align}
0 &= d_{20}+4d_{22}-4d_{23}+3d_{24}-22d_{25}-144d_{26}+4d_6+3d_{16}-10d_{17}-2d_{18}-72d_{19}
\end{align}

\end{document}